\documentclass[sigconf,natbib=true]{acmart} 

\usepackage[a-1b]{pdfx}

\usepackage{graphicx}
\usepackage{balance}
\usepackage{comment}
\usepackage{amsfonts}
\usepackage{color,soul}
\usepackage{hyperref}
\usepackage{subcaption}
\usepackage{xspace}
\usepackage{boxedminipage}
\usepackage{diagbox}
\usepackage{multirow}
\usepackage{paralist}
\usepackage{amsmath}
\usepackage{enumitem}
\usepackage{array}
\usepackage{mdframed}
\usepackage[normalem]{ulem}
\usepackage{tabularx}
\usepackage{pgf,tikz}
\usepackage{pgfplots}
\usepackage{mathrsfs}
\usepackage{diagbox}
\usepackage{footnote}
\usepackage{booktabs}
\usepackage{wrapfig}
\usepackage{url}
\usepackage{amsmath}
\usepackage{newtxmath}
\usepackage{arydshln}
\usepackage[flushleft]{threeparttable}

\usepackage[linesnumbered,ruled,vlined]{algorithm2e}

 \newcolumntype{L}[1]{>{\raggedright\arraybackslash}p{#1}}
\SetCommentSty{algcmmt}
\SetKwInput{KwInput}{Input}                
\SetKwInput{KwOutput}{Output}     

\DeclareMathOperator*{\argmax}{argmax}

\newcommand{\algname}{USTORY}

\AtBeginDocument{%
  \providecommand\BibTeX{{%
    \normalfont B\kern-0.5em{\scshape i\kern-0.25em b}\kern-0.8em\TeX}}}

\copyrightyear{2023}
\acmYear{2023}
\setcopyright{acmlicensed}\acmConference[SIGIR '23]{Proceedings of the 46th
International ACM SIGIR Conference on Research and Development in
Information Retrieval}{July 23--27, 2023}{Taipei, Taiwan}
\acmBooktitle{Proceedings of the 46th International ACM SIGIR Conference on
Research and Development in Information Retrieval (SIGIR '23), July 23--27,
2023, Taipei, Taiwan}
\acmPrice{15.00}
\acmDOI{10.1145/3539618.3591782}
\acmISBN{978-1-4503-9408-6/23/07}

\begin{document}

\title{Unsupervised Story Discovery from Continuous News Streams via Scalable Thematic Embedding}


\settopmatter{authorsperrow=4}

\author{Susik Yoon}
\affiliation{%
  \institution{UIUC}
}
\email{susik@illinois.edu}

\author{Dongha Lee}
\affiliation{%
  \institution{Yonsei University}
}
\email{donalee@yonsei.ac.kr}

\author{Yunyi Zhang}
\affiliation{%
  \institution{UIUC}
}
\email{yzhan238@illinois.edu}

\author{Jiawei Han}
\affiliation{%
  \institution{UIUC}
}
\email{hanj@illinois.edu}

\renewcommand{\shortauthors}{Susik Yoon, Dongha Lee, Yunyi Zhang, \& Jiawei Han}
\begin{abstract}
Unsupervised discovery of stories with correlated news articles in real-time helps people digest massive news streams without expensive human annotations. A common approach of the existing studies for unsupervised online story discovery is to represent news articles with symbolic- or graph-based embedding and incrementally cluster them into stories. Recent large language models are expected to improve the embedding further, but a straightforward adoption of the models by indiscriminately encoding all information in articles is ineffective to deal with text-rich and evolving news streams. In this work, we propose a novel \emph{thematic embedding} with an off-the-shelf pretrained sentence encoder to dynamically represent articles and stories by considering their shared temporal themes. To realize the idea for unsupervised online story discovery, a scalable framework \algname{} is introduced with two main techniques, \emph{theme- and time-aware dynamic embedding} and \emph{novelty-aware adaptive clustering}, fueled by lightweight story summaries. A thorough evaluation with real news data sets demonstrates that \algname{} achieves higher story discovery performances than baselines while being robust and scalable to various streaming settings. 
\end{abstract}



\maketitle

\section{Introduction}
\noindent\textbf{\uline{Background.}} Given a huge number of news articles (or simply \emph{article}s) being generated through digital mass media in real-time, a set of correlated articles naturally form a \emph{story} as they are consecutively published while describing relevant events under a unique theme. \emph{Unsupervised online story discovery} helps people digest massive news streams without human annotations by supporting real-time news curation services and downstream tasks such as recommendation, summarization, and event detection\,\cite{zhang2022evmine, news_recom, pdsum, arcus}.

To deal with \emph{text-rich} and \emph{evolving} properties of news streams, existing work for unsupervised online story discovery utilized keywords statistics to represent articles and incrementally cluster them into stories\,\cite{constream, miranda,staykovski,newslens}. As recent large language models (LLM) (e.g., BERT\,\cite{bert}) have shown strong performances in retrieval tasks\,\cite{tired, lotclass, IRLLM}, some works utilize LLMs for story discovery but only in an \emph{offline} and \emph{supervised} setting\,\cite{saravan, linger} with human-guided annotations which are expensive and easily outdated over time.

\noindent\textbf{\uline{Motivation.}}
In this work, for unsupervised online story discovery, we exploit off-the-shelf pretrained sentence encoders (PSE) based on LLMs and embed articles by using sentences as building blocks. This is a more practical and effective approach for dealing with text-rich and evolving news streams, because (1) a target corpus is dynamically changing, (2) true story labels are not readily available, and (3) modeling news articles with an LLM from scratch cause too high computational costs to support online applications.

\begin{figure}[t]
         \includegraphics[width=\columnwidth]{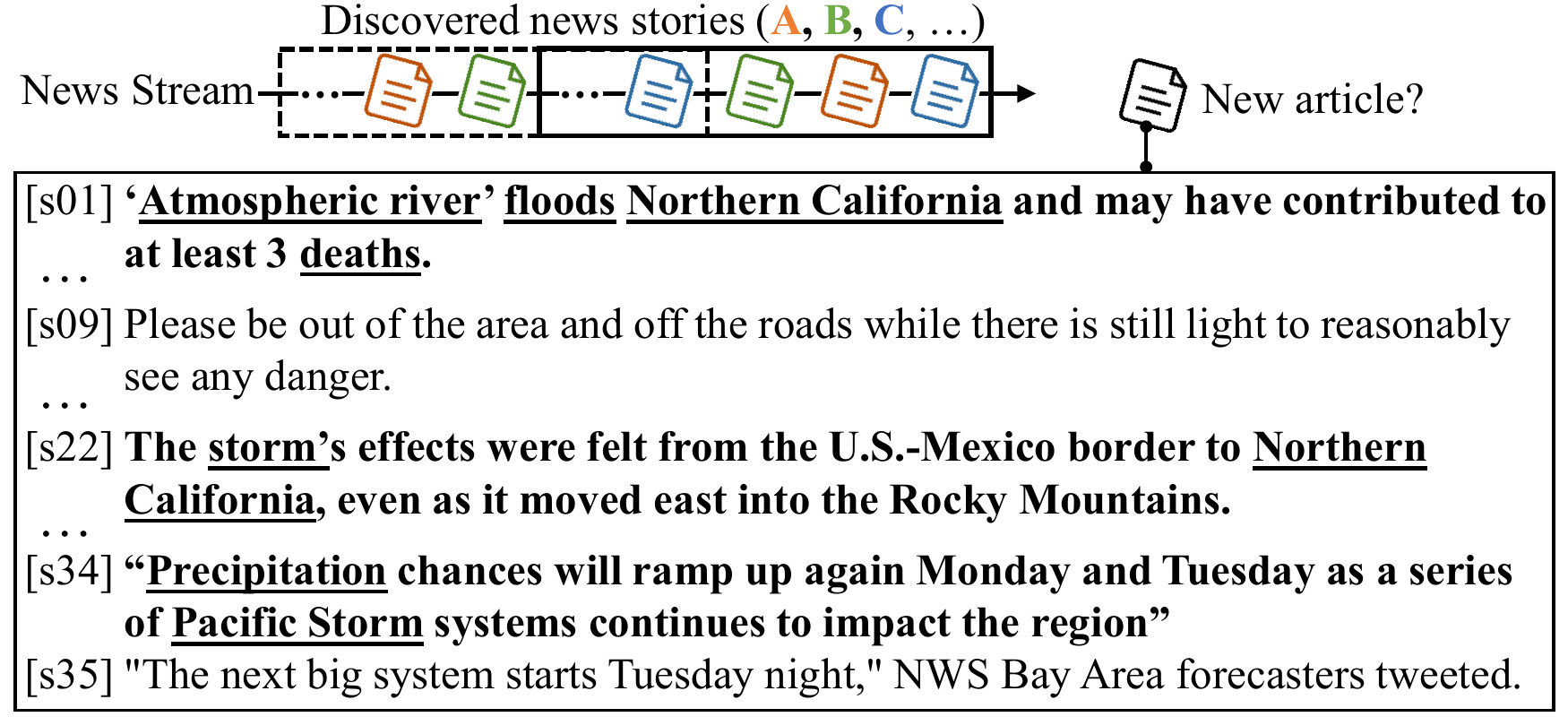}
         \vspace{-0.7cm}
         \caption{An article\,\cite{exart} about \textsf{2023 California Floods} where story-indicative sentences and keywords are highlighted.}
         \label{fig:motivation_ex_article}
         \vspace{-0.5cm}
\end{figure}


A straightforward way to embed an article with a PSE is to average individual sentence representations in the article into a single article representation, which we refer to as \emph{indiscriminative embedding}. While effective to some extent, compared with word- or article-level embedding with an LLM, it has clear limitations in considering the thematic contexts of news streams. For instance, in Figure \ref{fig:motivation_ex_article}, not all of the sentences equally contribute to representing the article for its story \textsf{2023 California Floods}. While some sentences (e.g., s1, s22, and s34) are highly relevant to the story as described with thematic keywords implying the story, the other sentences (e.g., s9 and s35) are just generic or too specific descriptions that are not necessarily relevant to the story. The indiscriminative embedding with a PSE, however, results in fixed and deterministic article representations regardless of such thematic contexts of news streams and thus fails to maximize the story discovery performance.


\begin{figure}[t]
        \includegraphics[width=\columnwidth]{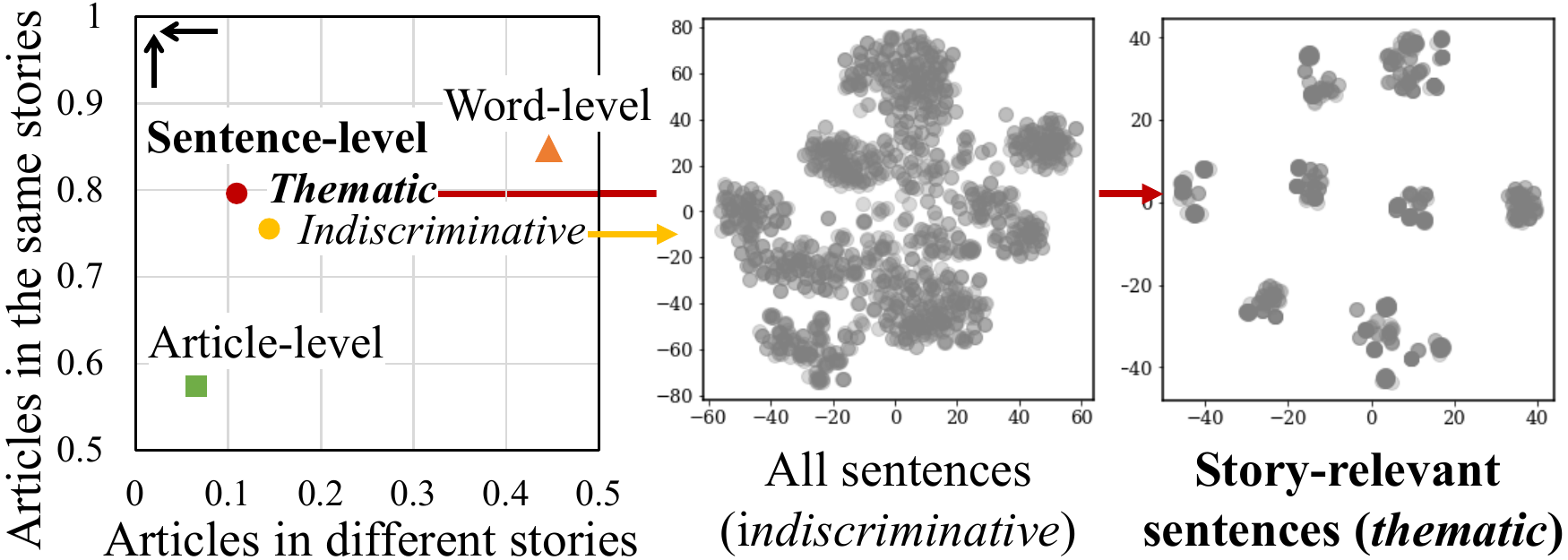}
        \vspace{-0.6cm}
        \caption{Left: Average cosine similarities of news articles\,\cite{WCEP} embedded by LLMs\,\cite{bert, sentencebert} with various granularity; higher (lower) is better for articles in the same (different) stories. Right: 2D-visualizations of sentences. } 
        \label{fig:motivation_embedding}
         \vspace{-0.6cm}
\end{figure}


\noindent\textbf{\uline{Main Idea and Challenges.}}
The key idea we employ to resolve this problem is \textbf{thematic embedding} with a PSE, which dynamically embeds articles and stories with their shared themes timely captured in the latest news streams. The temporal themes help focus on only story-relevant parts of articles for story discovery so that articles in the same story can be represented closer while being further from other stories. As demonstrated in Figure \ref{fig:motivation_embedding}, the thematic embedding identifies only story-relevant sentences which are more clearly clustered into distinct stories, and this naturally results in better article similarities than indiscriminative sentence-level embedding as well as word- or article-level embeddings.

However, implementing the idea of thematic embedding for unsupervised online story discovery pose considerable technical challenges: (1) unique themes of stories should be automatically identified (i.e., unsupervised) and timely updated (i.e., online) to derive up-to-date representations of articles and stories, (2) stories arbitrarily emerge and expire over time, so an adaptive mechanism is required to continuously cluster articles into stories, and (3) the embedding and clustering should be efficient and scalable to deal with massive news streams and support real-time applications. 


\noindent\textbf{\uline{Summary.}}
To this end, we propose a novel framework, \textbf{\algname{}}, for \uline{U}\! nsupervised \uline{Story} discovery with scalable thematic embedding. \algname{} can be instantiated with any existing PSEs and employs two main techniques: \emph{theme- and time-aware dynamic embedding} and \emph{novelty-aware adaptive clustering}; the former systematically identifies temporal themes from diverse contexts of news streams and embed articles and stories by considering their theme and time relevance (Section \ref{sec:semantic_embedding}). Then, the latter estimates article-story confidence scores to assign articles to existing stories or initiate novel stories. In the meantime, \algname{} manages only the minimum but sufficient summary of stories so as to guarantee single-pass processing of the assigned articles (Section \ref{sec:novelty_aware_clustering}).

We summarize the main contributions of this work as follows:
\vspace{-0.1cm}
\begin{itemize}[leftmargin=12pt, noitemsep]
\item  To the best of our knowledge, this is the first work to propose \emph{thematic embedding} with off-the-shelf pretrained sentence encoders for unsupervised online story discovery from news streams.
\item  We propose a framework \algname{}, implementing the idea of thematic embedding, which is \emph{scalable} with single-pass processing and \emph{compatible} with any pretrained sentence encoders. The source code is available at \url{https://github.com/cliveyn/USTORY}.
\item We demonstrate that \algname{} \emph{outperforms} existing baselines by up to 51.7\% in $B^3$-F1 and their PSE-variants by up to 12.5\% in $B^3$-F1 in three news data sets. The \emph{scalability} and \emph{robustness} of \algname{} are also verified through in-depth analyses.
\end{itemize}

\smallskip
\section{Related Work}
\label{sec:related_work}
Early work by Allan et al.\,\cite{TDT} introduced topic detection and tracking (TDT) concepts for organizing and mining news articles. News story discovery is one of the most widely studied relevant research topics in TDT, which is crucial for various downstream applications such as news recommendation\,\cite{news_recom}, summarization\,\cite{pdsum}, and fine-grained event mining\,\cite{zhang2022evmine}. Previous efforts in the news story discovery can be classified as retrospective (offline) story discovery or prospective (online) story discovery, where the former generates a story timeline\,\cite{retro2} or structure\,\cite{retro4} from a given set of articles usually for analyzing domain-specific events while the latter processes continuous and unbounded news streams to discover and track stories in real-time, which is the main scope of this paper.

For online news story discovery, clustering-based approaches have been widely used for embedded articles. One line of studies adopted a \emph{supervised} approach assuming some labeled articles and external knowledge (e.g., entity labels) are available to facilitate the embedding and clustering procedures. Specifically, labeled training data sets were used to learn an adjacency matrix of articles\,\cite{linger, storyforest, storyforest_conf} or to learn a similarity threshold for cluster assignment\,\cite{miranda, santos, saravan}. For instance, Miranda et al.\,\cite{miranda} introduced a multilingual clustering algorithm for news article streams, where an article-cluster similarity was measured by the weighted sum of TF-IDF sub-vectors similarities and temporal similarities modeled in a Gaussian distribution. The weights for the various types of similarities were learned from labeled training data sets. Some work fine-tuned an LLM with the labeled training data sets for embedding articles along with external entity knowledge\,\cite{saravan, linger}. 

However, in an online setting, the labeled articles are rarely available and quickly outdated. The other line of studies thus tried to embed and cluster articles in an \emph{unsupervised} manner. ConStream\,\cite{constream} is one of the popular clustering algorithms for document streams and has been widely used as a competitive baseline. It managed keyword frequencies of articles and clusters them with a notion of micro-cluster. A recent work NewsLens\,\cite{newslens} found overlapping keywords of articles to build a local topic graph and applied a community detection algorithm to it to detect stories. Staykovski et al.\,\cite{staykovski} improved NewsLens further by using the sparse TF-IDF embedding (which outperformed the dense doc2vec\,\cite{doc2vec} embedding alternatively proposed in the work) of articles to build a local topic graph. The existing methods, however, focus on explicit keywords statistics of articles, limiting the capability to capture implicit local contexts inside the articles. Furthermore, their fixed article representations do not fully capture evolving global contexts of news streams. In this work, we capture the local contexts of articles with a PSE, while dynamically exploiting it through thematic embedding to adapt to the global contexts of news streams.

Besides, some studies\,\cite{newsembed, newsbert} proposed \emph{offline} and \emph{supervised} fine-tuning of language models to specifically model news articles with relevant labeled data sets and tasks. Such fine-tuned models inherently output a fixed and deterministic representation of an input article, i.e., indiscriminative embedding, without considering the global contexts of news streams. However, their pretrained models can also be equipped in our framework for unsupervised online story discovery by being dynamically exploited over news streams through thematic embedding.



\section{Problem Setting}
Let $a = [s_1, s_2, \ldots, s_{|a|}]$ be a news article\,(or simply \emph{article}) composed of sentences $s$ and $\mathcal{C} = [a_1, a_2, \ldots, a_{|\mathcal{C}|}]$ be a news story\,(or simply \emph{story}) composed of correlated articles with a unique theme, such as \textsf{California Floods}, and consecutively published for a certain duration of time. We assume that each article belongs to a single story, and each story has at least $M$ articles. A news stream $\mathcal{A} (= \ldots, a_{i-1}, a_{i}, a_{i+1}, \ldots)$ is a continuous and unbounded sequence of articles consecutively published at a timestamp $t_{a_i}$. A \emph{sliding window} $\mathcal{W}$ of size $W$ slid by $S$ determines a context of the latest articles and ongoing stories in $\mathcal{A}$. We set $W=7$ days and $S=1$ day by default (i.e., if no articles are added to a story for a week, it is considered expired), while they can be alternatively set as the number of articles. Then, Definition \ref{def:UOSD} gives a formal definition of the problem considered in this work.

\vspace{-0.1cm}
\begin{definition}
\label{def:UOSD} 
({\sc Unsupervised Online Story Discovery}) Given a news stream $\mathcal{A}$, the unsupervised online story discovery is to incrementally update a set $\mathbb{C}_{\mathcal{W}}$ of stories from the articles in every sliding window $\mathcal{W}$ without any human supervision or story labels.
\end{definition}
\vspace{-0.1cm}


\section{Thematic Embedding}
\label{sec:semantic_embedding}
\subsection{Motivation}
The efforts to encode texts have been long made from symbolic-based models (e.g., bag-of-words) to recent LLMs\,\cite{bert, roberta, T5}. As demonstrated in Figure \ref{fig:motivation_embedding}, using individual word-level or entire article-level granularity for embedding an article with an LLM may not be optimal since they are either too fine-grained (former) or too coarse-grained (latter) to get its specific story-indicative semantics, which is shared within the same story but not in different stories. Exploiting sentence-level information, on the other hand, effectively balances the abstractness of semantics and naturally meets the input constraints for typical LLMs (e.g., 512 tokens). Recent PSEs, that fine-tune LLMs with benchmark data sets and tasks for specifically embedding sentences, have shown state-of-the-art sentence embedding capabilities across various domains\,\cite{sentencebert,sentencet5}. In this work, we exploit off-the-shelf PSEs by regarding sentences as building blocks for embedding articles. 

Exploiting a PSE to embed a long article (i.e., tens of sentences) gives various design choices; for instance, concatenating or mean-pooling of sentences can be straightforward alternatives\footnote{Refer to Section \ref{exp:alternatives} for the comparison of different alternative embedding strategies.}. PSEs can effectively capture the local semantics of individual sentences in an article, but as shown in Figure \ref{fig:motivation_ex_article}, the sentences are not necessarily relevant to the theme of the article's story. Thus, \algname{} employs \emph{thematic embedding} with a PSE, which first identifies the temporal themes of articles given a particular context of news streams (Section \ref{sec:temp_theme_iden})  and then dynamically embeds the articles and stories considering their theme- and time-relevance (Section \ref{sec:article_story_embedding}).

\subsection{Temporal Theme Identification} 
\label{sec:temp_theme_iden}
Let \emph{corpus} be a set of articles collected over a period of time and \emph{corpora} be a set of the corpus. Then, each corpus must have a \emph{temporal theme} that uniquely represents the corpus especially in the latest temporal context of corpora. We model the temporal theme through a keywords retrieval process because it is efficient to mine keywords with a simple tokenization pipeline and it is effective to represent a theme explicitly with a diverse combination of keywords. Specifically, we identify thematic keywords of a corpus in context corpora by conjunctively considering (1) \emph{recency}, (2) \emph{popularity}, and (3) \emph{distinctiveness} of keywords.

For example, let a context corpora $\mathbb{D}$ be the latest articles in news streams and a target corpus $d$ be the articles about \textsf{2023 California Floods} in $\mathbb{D}$. The keywords such as \emph{`Northern California'}, \emph{`Evacuation order'}, \emph{`Pacific Storm'}, \emph{`Die'}, and \emph{`Rescue'} may collectively describe the temporal theme of $d$, while their compositions and importances will change over time. Suppose that the term \emph{`Pacific Storm'} has consistently appeared in $d$ for the last 7 days by $L$ times in each day, while \emph{`Evacuation order'} has appeared more actively for the last 3 days by $2L$ times in each day. 
While \emph{`Evacuation order'} was less popular than \emph{`Pacific Storm'} for the last week (i.e., $2L\!\times\!3$ days < $L\!\times\!7$ days), it would be more valuable to represent the recent theme of $d$. At the same time, other recent popular keywords such as \emph{`Die'} or \emph{`Rescue'} would be devalued if they are also frequently observed in another corpus in $\mathbb{D}$ (e.g., \textsf{Russia-Ukraine Conflict}).


To this end, we naturally incorporate \emph{time-decaying property} into popular ranking functions for discriminative information retrieval (e.g., TF-IDF\,\cite{tfidf} and BM25L\,\cite{bm25l})\footnote{Any frequency-based function can be adopted with the time-decaying property. We used TF-IDF as default which showed better results than the other functions.} to identify thematic keywords. 

\begin{definition}
\label{def:keywords} 
({\sc Thematic Keywords}) Given a target corpus $d$ in a context corpora $\mathbb{D}$, a set $\mathcal{K}_{d}$ of the top $N$ thematic keywords that best describe the temporal theme of $d$ at a time $t_{c}$ is
\begin{align*}
\small
\begin{split}
\mathcal{K}_{d} = \{(k_1,w_{k_1}), (k_2,w_{k_2}), \ldots, (k_N,w_{k_N})\}, \text{where}
\end{split}
\end{align*}
\vspace{-0.3cm}
\begin{equation}
\small
\label{eq:keywords_importance} 
\begin{split}
w_{k} &= rec\text{-}pop(k,d, t_c)\cdot dist(k,\mathbb{D}) \\
&= \sum_{k^j \in \mathcal{T}_{d}} \exp(-\frac{|t_{k^j}-t_{c}|}{\delta_k}) \cdot \log(\frac{|d_i \in \mathbb{D}|+1}{|d_i \in \mathbb{D} : k \in \mathcal{T}_{d_i}|+1}+1),
\end{split}
\end{equation}
where $k_i$ is a single- or multi-token term appearing in $d$ ranked by its importance $w_{k_i}$ and $\mathcal{T}_d$ is a set of all term appearances in $d$. 
\end{definition}

The score function $rec\text{-}pop(k,d, t_c)$ is time-decaying term frequency, where each term appearance $k^j$ at time $t_{k^j}$ is counted (for \emph{popularity}) while being exponentially decayed by its temporal difference from $t_c$ (for \emph{recency}). The score function $dist(k,\mathbb{D})$ is inverse corpus frequency to measure how unique $k$ is in $\mathbb{D}$ (for \emph{distinctiveness}). The decaying factor $\delta_k$ controls the degree of decaying and can be set to the total time span of $\mathbb{D}$ (i.e., $\delta_k =max(t_{k^j}\!)\!-\!min(t_{k^j})\!+\!1 \text{ for } k^j \! \in \! \mathcal{T}_{\mathbb{D}}$). 

\subsection{Theme/Time-aware Dynamic Embedding}
\label{sec:article_story_embedding}

\textbf{\uline{Article Embedding.} }
A temporal theme can act as key guidance in embedding an article with a PSE; the article is best represented with the theme by focusing on only the theme-relevant parts of the article. Given a certain temporal theme, as a form of thematic keywords, we dynamically represent an article by pooling the representations of sentences in the article weighted by their \emph{theme relevance}, which considers both the frequency and the importance of the thematic keywords found in each sentence. 

\begin{definition}
\label{def:article_emb} 
({\sc Article Representation}) Given a thematic keywords set $\mathcal{K}_{d}$, derived from a target corpus $d$ in context corpora $\mathbb{D}$, a representation $E_{a|d}$ of a target article $a$ given $d$ is
\begin{equation}
\begin{split}
E_{a|d} = \sum_{s_l \in a}\frac{\sum_{k_i \in \mathcal{K}_{d}}{|k_i^j  \in  \mathcal{T}_{s_l}|w_{k_i}}}{\sum_{k_i \in \mathcal{K}_{d}}\!|k_i^j   \in  \mathcal{T}_{a}|w_{k_i}}enc(s_l),
\end{split}
\end{equation}
where $\mathcal{T}_{a}$ and $\mathcal{T}_{s}$ are the term appearance sets of $a$ and its sentence $s$, respectively, and $enc(s)$ is a representation of $s$ by a PSE.
\end{definition}

\begin{figure}
    \centering
    \includegraphics[width=\columnwidth]{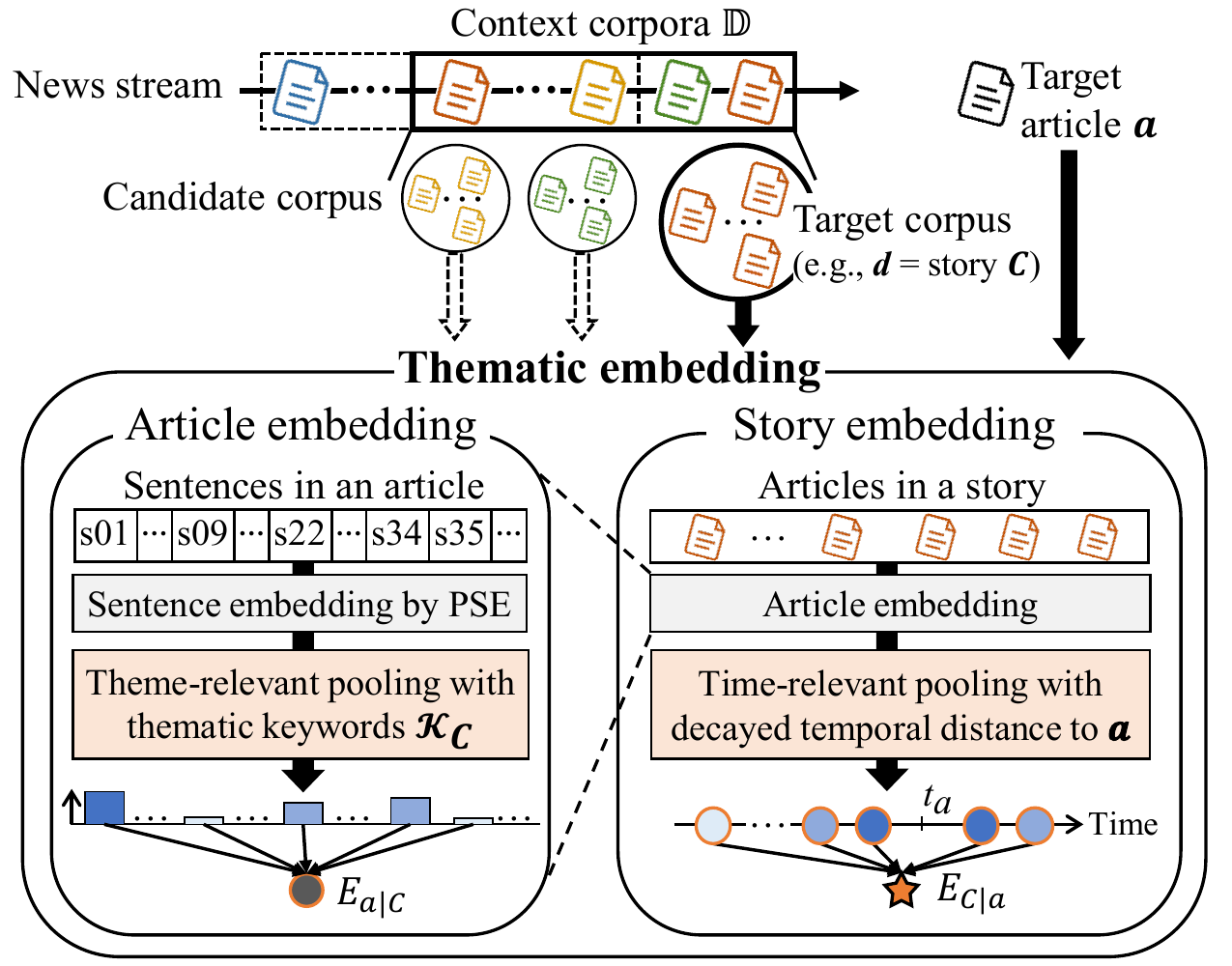}
    \vspace{-0.7cm}
    \caption{Thematic embedding with an article in Figure \ref{fig:motivation_ex_article} and its story \emph{2023 California Floods} as the target article and story.}
    \vspace{-0.4cm}
    \label{fig:dynamic_embedding}
\end{figure}

\noindent\textbf{\uline{Story Embedding.} }
As a story is basically a cluster of articles, a typical way to represent it is to average all the incorporated article representations (i.e., a cluster center). However, such a static story embedding does not correctly capture the temporal theme of the story, which gradually evolves with newly added articles. Thus, we dynamically represent a story given a target article (i.e., at a specific time of the article) by pooling the representations of articles in the story weighted by their \emph{time relevance} to the target article.
\vspace{-0.1cm}
\begin{definition}
\label{def:story_emb} 
({\sc Story Representation}) A representation $E_{\mathcal{C}|a}$ of a target story $\mathcal{C}$ given a target article $a$ is
\vspace{-0.1cm}
\begin{equation}
\begin{split}
E_{\mathcal{C}|a} = \sum_{a_i \in \mathcal{C}}\frac{ \exp(-|t_a-t_{a_i}|/\delta_C)}{\sum_{a_j \in \mathcal{C}} \exp(-|t_a-t_{a_j}|/\delta_C)}E_{a_i|C},
\end{split}
\vspace{-0.1cm}
\end{equation}
where the time-decaying property is applied to the temporal distance and the decaying factor $\delta_{C}$ can be set to the total time span of the story (i.e., $\delta_C = max(t_{a_i}) - min(t_{a_i})$ for $\forall a_i \in C$).
\end{definition}
\vspace{-0.1cm}

Figure \ref{fig:dynamic_embedding} illustrates an example of thematic embedding. Suppose that an article in Figure \ref{fig:motivation_ex_article} is a target article $a$, and a story $C$ of articles about \textsf{2023 California Floods} is a target corpus $d$. Then, the thematic keywords set $\mathcal{K}_{C}$ is identified as a temporal theme of $C$. When embedding $a$ given $C$, the theme-relevant sentences (e.g., s01, s22, and s34) served as key ingredients for representing $a$. At the same time, when embedding $C$ given $a$, the articles in $C$ temporally close to $a$ contribute more to represent $C$. Note that articles and stories can be dynamically represented depending on the target articles and stories, facilitating more effective story discovery.

\begin{figure*}
    \centering
    \includegraphics[width=0.83\paperwidth]{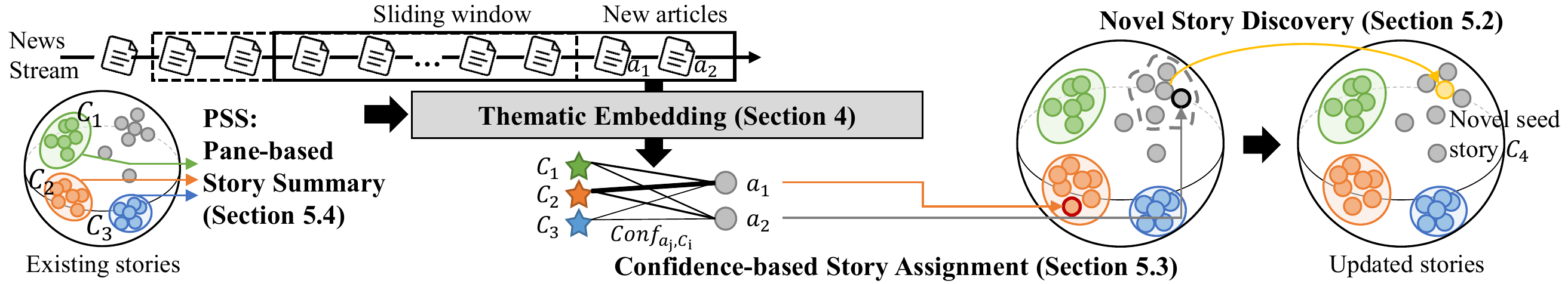}
    \vspace{-0.5cm}
    \caption{The overall procedure of \algname{}.}
    \vspace{-0.3cm}
    \label{fig:method_overview}
\end{figure*}

\section{Novelty-aware Adaptive Clustering}
\label{sec:novelty_aware_clustering}
\begin{algorithm}[!t]
    \label{alg:overall}
    \caption{Overall Procedure of \algname{}}
    \small
    \DontPrintSemicolon
    \KwInput{a news stream $\mathcal{A}$, a pre-trained sentence encoder $enc$}
    \KwOutput{a set $\mathbb{C}_{\mathcal{W}}$ of stories in every sliding window $\mathcal{W}$}
    \For{every sliding window $\mathcal{W}$ from $\mathcal{A}$}
    {
        \For{new articles $a_j$ in $\mathcal{W}$}
        {
            $[enc(s)|s \in a_j] \leftarrow$ sentence representations of $a_j$\;
        }
        \If{$\mathbb{C}_{\mathcal{W}}$ is empty}
        {
                \tcc{Novel (initial) Story Discovery (Section \ref{sec:seed_story})}
        $\mathbb{C}_{\mathcal{W}} \leftarrow$ initialize seed stories from new articles;
        }
            
        \tcc{Confidence-based Story Assignment (Section \ref{sec:conf_story_assign})}
        \For{unassigned article $a_j$ in $\mathcal{W}$}
        {
            $C_i^* \leftarrow \argmax\limits_{C_i \in \mathbb{C}_{\mathcal{W}}}conf_{a_j|C_i}$  \tcp*{The most confident story}
            \If{$conf_{a_j|C_i^*} \geq \gamma $}
            {
                Assign $a_j$ to $C_i^*$\\
                Update $PSS_{C_i^*}$ \tcp*{Story summary (Section \ref{sec:story_summary})}
            }
        }
        
        \tcc{Novel Story Discovery (Section \ref{sec:seed_story})}
        $\mathbb{C}_{\mathcal{W}} \leftarrow$ add seed stories from unassigned articles;
        
        Return $\mathbb{C}_{\mathcal{W}}$\;
    }
\end{algorithm}

\subsection{Overview}
Using the thematic embedding in Section \ref{sec:semantic_embedding}, \algname{} incrementally clusters articles into stories, while adaptively discovering novel stories. The overall procedure of \algname{} is illustrated in Figure \ref{fig:method_overview} and outlined in Algorithm 1. For every sliding window of a news stream, \algname{} gets sentence representations of new articles from a PSE (Lines 1-3) and conducts two steps. First, if there are no existing stories in the current window, as \emph{novel story discovery} step, new articles are represented with their own theme, and seed stories are identified by cluster center initialization (Lines 4-5). Then, as \emph{confidence-based story assignment} step (Lines 6-10), unassigned articles (including the new articles) are inspected if they can be confidently added to one of the existing stories. Each article-story pair is dynamically embedded and their confidence scores are derived from their thematic similarity. An article is assigned to the most confident story if the score exceeds a threshold. In the meanwhile, a summary of existing stories is utilized and updated. Finally, the novel story discovery step is conducted for the remaining unassigned articles to form new stories (Line 11) and all the discovered current stories are reported (Line 12). The two steps are detailed in the following sections.

\subsection{Novel Story Discovery}
\label{sec:seed_story}
\textbf{\uline{Initial Article Embedding.} }
When there are no existing stories or confident stories to be assigned to, unassigned articles (e.g., newly published or previously unassigned) are used to find novel seed stories. Since there are no existing themes or adequate themes to be considered, the unique theme of each article itself is identified for the thematic embedding of the article. Specifically, in Definition \ref{def:keywords}, the context corpora $\mathbb{D}$ becomes all articles in a sliding window $\mathcal{W}$ and the target corpus $d$ becomes a target article $a$. Then, the \emph{article-indicative} thematic keywords set $\mathcal{K}_{a}$ are derived and it leads to the representation $E_{a|\{a\}}$ of the article $a$ by Definition \ref{def:article_emb}.

\noindent\textbf{\uline{Seed Stories Discovery.} }
Once the articles are embedded to better reveal their own themes, the articles under similar themes are more likely to be closer than those under different themes. Thus, a typical cluster center initialization technique can be applied to the thematically embedded initial representations of unassigned articles to find novel seed stories. Motivated by the popularly used k-means++ initialization\,\cite{kmeans++}, \algname{} finds the seed centers with the lowest inertia (i.e., the sum of cosine similarities between each article and its centers) to get the most thematically distinctive seed stories. The number of seeds can be decided by dividing the number of unassigned articles by the minimum number $M$ of articles to initiate a story, which is dependent on user preferences and application requirements\footnote{Typically, a news story has 5 to 20 articles on its first day in the real data sets\,\cite{miranda,WCEP}.}. If there are no such constraints available, an existing data-driven approach for deciding the number of clusters such as LOG-Means\,\cite{log-means} or Silhouette Coefficient\,\cite{silhouettes} can be applied.

\subsection{Confidence-based Story Assignment}
\label{sec:conf_story_assign}
When there are existing stories, each unassigned article is evaluated to be assigned to one of the stories. Specifically, the unique temporal themes of existing stories are updated, and then the thematic similarity between each pair of stories and articles is estimated to get the robust article-story confidence score.  

\noindent\textbf{\uline{Thematic Embedding.}} A \emph{story-indicative} thematic keywords set $\mathcal{K}_{C}$ is derived by Definition \ref{def:keywords}, by setting the context corpora $\mathbb{D}$ to a set $\mathbb{C}_{\mathcal{W}}$ of stories in a sliding window $\mathcal{W}$ and the target corpus $d$ to the articles in a story $C$. Then, for each pair of article $a$ and story $C$, the article representation $E_{a|C}$ and the story representation $E_{C|a}$ are derived by Definition $\ref{def:article_emb}$ and \ref{def:story_emb}, respectively.


\noindent\textbf{\uline{Article-Story Thematic Similarity.}}
Given a pair of articles and stories, \algname{} quantifies their \emph{thematic similarity} by conjunctively considering their semantic themes and symbolic themes. In brief, the former is estimated by the cosine similarity between their thematic representations, and the latter is estimated by the divergence of their thematic keyword distributions. These two types of similarities complement each other to estimate more robust thematic similarities. 

For instance, some articles about \textsf{2023 California Floods} might have a few sentences of slightly different thematic semantics according to the writers' perspectives (e.g., one may focus on the rescue while another may focus more on the victim), but the overall thematic keywords distributions of these articles would be similar as they are describing the same event. On the other hand, some articles about \textsf{2023 California Floods} and the other articles about \textsf{Russia-Ukraine Conflict} might happen to have a few sentences of similar thematic semantics describing casualties, but their overall thematic keywords distributions would be different (e.g., for articles in \textsf{2023 California Floods} \emph{`die'} is co-occurred more frequently with \emph{`floods'} or \emph{`storm'}). Definition \ref{def:thematic_similarity} formally formulates the thematic similarity between an article and a story.

\vspace{-0.1cm}
\begin{definition}
\label{def:thematic_similarity} 
({\sc Thematic Similarity}) A thematic similarity between an article $a$ and a story $C$ is calculated as 
\vspace{-0.1cm}
\begin{equation}
\small
\begin{split}
    sim_{theme}(a,C) = max(0,\text{cos}(E_{a|C}, E_{C|a})) \cdot JSD(P_{a,\mathcal{K}_C}\|P_{C,\mathcal{K}_C}).
\end{split}
\vspace{-0.1cm}
\end{equation}
The first term is a cosine similarity, with the negative values truncated to zero, between the thematic representations of $a$ and $C$. The second term is the JS-divergence\,\cite{JSD}\footnote{Among popular divergence measures\,\cite{JSD,KLD,EMD}, we chose JS-divergence\,\cite{JSD} since it is bounded within a finite interval $\in [0, 1]$, can be defined even when a keyword exists only in a story, and has a low time complexity, i.e., $O(|\mathcal{K}_{C}|)$.} between the thematic keyword probability distributions of $a$ and $C$. The keyword probability $P(k_i|a,\mathcal{K}_C)$ in $P_{a,\mathcal{K}_C}$ is estimated as $\frac{|k_i \in \mathcal{T}_a|}{\sum_{k_j \in \mathcal{K}_C}|k_j \in \mathcal{T}_a|}$ where $\mathcal{T}_a$ is the term appearance set of $a$ (and similarly for $P(k_i|C,\mathcal{K}_C)$ in $P_{C,\mathcal{K}_C}$).
\end{definition}
\vspace{-0.1cm}

\noindent\textbf{\uline{Article-Story Assignment.}} Finally, \algname{} derives a confidence score $conf_{a,C} \in [0,1]$ for an article $a$ to be assigned to a story $C$ by comparing their thematic similarity with the other thematic similarities of all possible candidate assignments.

\begin{definition}
\label{def:story_confidence} 
({\sc Article-Story Confidence}) Given a target article $a$ and a set of candidate stories $C_i \in \mathbb{C}_{\mathcal{W}}$, the article-story confidence score for $a$ to be assigned to $C_i$ is
\begin{equation}
\small
conf_{a,C_i} = \frac{\exp(T \cdot sim_{theme}(a,C_i))}{\sum_{C_j \in \mathbb{C}_\mathcal{W}}\exp(T \cdot sim_{theme}(a,C_j))},
\end{equation}
where $T$ is a temperature for scaling the score distribution.

\end{definition}
Then, the article is assigned to the story with the highest confidence score if it exceeds the threshold $\gamma = 1- (1-1/|\mathbb{C}_\mathcal{W}|)^T$, indicating an adjusted random assignment probability\,\cite{taxocom}, or otherwise remains unassigned\footnote{The sensitivity analysis and the guidance on $T$ are given in Section \ref{sec:hyperparameter}}. The unassigned articles are used to find seed stories as introduced in Section \ref{sec:seed_story} and repeatedly inspected to be assigned to the updated stories in later sliding windows.

\subsection{Scalable Processing with Story Summary}
\label{sec:story_summary}

\noindent\textbf{\uline{Story Summary.}} Identifying temporal themes and dynamically embedding articles and stories from scratch in every sliding window cause considerable computation overheads, which is not practical in an online scenario. To realize scalable online processing, \algname{} uses a novel data structure, called \emph{pane-based story summary}, motivated by the pane-based aggregation\,\cite{pane} that has been widely adopted for efficient and scalable stream algorithms\,\cite{nets, pane_ex1, stare, mdual}.

\vspace{-0.1cm}
\begin{definition}
\label{def:PSS} 
({\sc Pane-based Story Summary (PSS)}) Let panes $p_i$ be non-overlapping subsets of consequent articles in a news stream and story panes $p_i^C$ of a story $C$ be a set of articles in both $C$ and $p_i$. Then, a pane-based story summary of $C$,
\vspace{-0.1cm}
\begin{equation}
\small
PSS_C = \{ \langle p_i: \langle \: |p_i^C|,\: tf(\cdot, p_i^C),\: \Sigma_{a_j \in p_i^C} E_{a_j|C}\: \rangle \rangle \},
\vspace{-0.1cm}
\end{equation}
maps a pane to the triplet of the number of articles, the term frequencies, and the sum of article representations in $p_i^C$.
\vspace{-0.1cm}
\end{definition}

The size of a pane determines the granularity of the story summary. While any common divisor of the window size and the slide size can be used, we set the pane size to be the slide size (i.e., the greatest common divisor) to maximize efficiency. In the confidence-based story assignment step in every sliding window, \algname{} uses PSS to identify temporal themes and derive dynamic article and story representations, without accessing all previous articles in stories. Note that \algname{} updates the triplet in PSS in an \emph{additive} manner whenever articles in stories are updated as the window slides. The sufficiency of PSS and the complexity of \algname{} realized by utilizing PSS are analyzed as follows.


\noindent\textbf{\uline{Efficiency Analysis.}} With the help of PSS, \algname{} guarantees \emph{single-pass processing} on the assigned articles; once an article is assigned to a story it can be discarded and only PSS is used for the following procedures. Theorem \ref{theorem:sufficiency} and Theorem \ref{theorem:complexity} respectively proves the sufficiency of PSS for the story assignment and shows the time and space complexities of \algname{} when utilizing PSS.

\begin{theorem}
\vspace{-0.1cm}
\label{theorem:sufficiency} 
{\sc (Sufficiency of PSS)} The confidence-based story assignment step requires to identify thematic keywords $\mathcal{K}_{C}$ and a story representation $E_{C|a}$ for each story $C$ and each unassigned article $a$ (note that $E_{a|C}$ is directly derived from $\mathcal{K}_{C}$ and $a$). A pane-based story summary $PSS_C$ is sufficient for deriving $\mathcal{K}_{C}$ and $E_{C|a}$.
\end{theorem}
\vspace{-0.3cm}
\begin{proof}
$\mathcal{K}_{C}$ and $E_{C|a}$ can be accurately derived by only using the triplet information in $PSS_C$. First, by Definition \ref{def:keywords}, the importance $w_k$ of a thematic keyword $k$ with a story $C$ in a sliding window $\mathcal{W}$ at a current timestamp $t_c$ is computed from $rec\text{-}pop(k,C,t_c)$ and $dist(k,\mathbb{C}_{\mathcal{W}})$. Since a pane $p_i$ represents the articles in the same slide in $\mathcal{W}$, i.e., $\forall \! a \! \in \! p_i \!:\! t_a \! = \! t_{p_i}$, $rec\text{-}pop(k,C,t_c)$ and $dist(k,\mathbb{C}_{\mathcal{W}})$ are calculated with the term frequency $tf(k,p_i^C)$ in $PSS_C$.
\vspace{-0.1cm}
\begin{equation}
\small
\begin{split}
    rec\text{-}pop(k,C,t_c) 
    &= \sum_{p_i \in \mathcal{W}} \Big( \exp(- \frac{|t_{p_i}\! - \! t_c|}{\delta_k}) \cdot tf(k,p_i^C) \Big) \text{ and} \\
    dist(k,\mathbb{C}_{\mathcal{W}}) 
    &= \log(\frac{|\mathbb{C}_{\mathcal{W}}| \! + \! 1}{\sum \vmathbb{1}_{\sum_{p_i \in \mathcal{W}}tf(k,p_i^C)>0}  \! + \! 1} \! + \! 1).
\end{split}
\end{equation}
Then, in Definition \ref{def:story_emb}, $E_{C|a}$ can be reformulated as the sum of time-decaying article representations divided by the time-decaying count of articles. Thus, it can be calculated with the articles count $|p_i^C|$ and the article representations sum $\sum_{a_j \in p_i^C}E_{a_j|C}$ in $PSS_C$. 
\begin{equation}
\small
\begin{split}
    E_{C|a} 
    &= \frac{\sum_{p_i \in \mathcal{W}} \big( \exp(-|t_a - t_{p_i}| / \delta_C) \cdot \sum_{a_j \in p_i^C}E_{a_j|C} \big) }{\sum_{p_i \in \mathcal{W}} \big( \exp(-|t_a - t_{p_i}| / \delta_C) \cdot |p_i^C| \big) }.
\end{split}   
\end{equation}
\vspace{-0.1cm}
\end{proof}
\vspace{-0.3cm}
\begin{theorem}
\vspace{-0.1cm}
\label{theorem:complexity} 
{\sc (Complexities of \algname{})} Let $N_W$ and $N_S$ be the numbers of articles in a window and a slide, respectively, and $N_C$ be the number of existing stories. Then, the time and space complexities of \algname{} are $O(N_CN_W+ N_S)$ and $O(N_W + \frac{N_W}{N_S} N_C)$, respectively.  
\end{theorem}
\vspace{-0.3cm}
\begin{proof}
The time complexity of \algname{} is specifically divided into that for each step: $O(N_S)$ for embedding articles in every sliding window with a PSE, $O(N_CN_W)$ for the novelty story discovery step, $O(N_CN_W)$ for the confidence-based story assignment step, and $O(N_S)$ for updating PSS. Thus, the total time complexity of \algname{} is $O(N_CN_W+N_S)$.
Similarly, the space complexity of \algname{} is specifically divided into that for managing stories and articles: $O(\frac{N_W}{N_S} N_C)$ for managing PSS where $\frac{N_W}{N_S}$ is the number of panes and $O(N_W)$ for managing articles. Thus, the total space complexity of \algname{} is $O(N_W + \frac{N_W}{N_S} N_C)$.
\end{proof}
\vspace{-0.1cm}

Since $N_C, N_S \ll N_W$ in practice, the time and space complexities of \algname{} are \emph{linear} to $N_W$, mainly affected by a window size $W$.

\section{Experiments}
We conducted extensive experiments to evaluate the performance of \algname{}, of which results are summarized as follows.

\begin{itemize}[leftmargin=10pt, noitemsep]
    \item \algname{} outperformed the existing unsupervised online story discovery algorithms and their variants with a PSE in benchmark news data sets in terms of $B^3$-F1, AMI, and ARI (Section \ref{sec:clustering_evaluation}).
    
    \item The main idea employed in \algname{} was demonstrated to be effective through ablation studies on the theme- and time-aware components (Section \ref{sec:clustering_evaluation}) and comparison with alternative embedding strategies (Section \ref{exp:alternatives}).     
    \item \algname{} was scalable to the variation of sliding window sizes and robust to hyperparameters (Section \ref{sec:hyperparameter}). 
    \item \algname{} discovered more quality embedding space and news stories than baselines from a real news stream (Section \ref{sec:casestudy}).
\end{itemize}
\begin{table}[!t]
\caption{Data sets.}
\vspace{-0.3cm}
\small
\label{tbl:datasets}
\begin{tabular}{cccc}
\toprule
Data set                 & \begin{tabular}[c]{@{}c@{}}\#Articles\\ {\footnotesize (Avg \#Sentences)}\end{tabular}& \begin{tabular}[c]{@{}c@{}}\#Stories \\ {\footnotesize (Avg \#Articles/day)} \end{tabular}\\ \hline
Newsfeed\,\cite{miranda, newsfeed} & 16,136 (21.4)                & 788 (8)       \\
WCEP18\,\cite{WCEP}     & 47,038 (26.9)               & 828 (18)      \\
WCEP19\,\cite{WCEP}     & 29,931 (27.6)                & 519 (18)      \\
USNews (case study)     & 2,744 (34.9)                & -      \\
\bottomrule
\end{tabular}
\vspace{-0.5cm}
\end{table}

\subsection{Experiment Setting}
\label{exp:setting}
\noindent\textbf{\uline{News Streams.} } We used four real news data sets for evaluation, summarized in Table \ref{tbl:datasets}. Newsfeed\,\cite{miranda, newsfeed} is a multilingual news data set collected from Newsfeed Service\,\cite{newsfeed_service} in 2014, and we used English news articles with story labels. WCEP\,\cite{WCEP} is a benchmark news data set collected from Wikipedia Current Event Portal and the Common Crawl archive. We used articles in the stories of at least 50 articles and published in 2018 and 2019 (i.e., WCEP18 and WCEP19). For a qualitative case study, we prepared USNews by collecting news articles through NewsAPI\,\cite{newsapi} with a query of `United States' for a month. Each data set is simulated as a news stream by feeding articles into sliding windows in chronological order. The window size $W$ and the slide size $S$ of the sliding window are set to 7 days and a day, respectively.


\bgroup
\def\arraystretch{0.9}%
\begin{table*}[!t]
    \caption{Performance comparison results (the highest scores are highlighted in bold).}
    \vspace{-0.3cm}
    \label{tbl:overall_accuracy}
    \begin{threeparttable}
    \begin{tabular}{c|l|ccc|ccc|ccc}
\toprule
\multicolumn{2}{c|}{}                                                             & \multicolumn{3}{c|}{Newsfeed}     & \multicolumn{3}{c|}{WCEP18}     & \multicolumn{3}{c}{WCEP19}            \\
\multicolumn{2}{c|}{}                                & $B^{3}$-F1                     & AMI                                      & ARI                             & $B^{3}$-F1                     & AMI                            & ARI                             & $B^{3}$-F1                      & AMI                            & ARI   \\ \hline
\multirow{5}{*}{
\begin{tabular}[c]{@{}c@{}} Baselines \\ (ordered by the \\average $B^3$-F1) \end{tabular}}                                                       & ConStream                      & 0.314                          & 0.128                                    & 0.069                           & 0.408                          & 0.444                          & 0.222                           & 0.400                           & 0.497                          & 0.292                                      \\
                                                                                 & NewsLens                       & 0.481                          & 0.309                                    & 0.077                           & 0.527                          & 0.490                          & 0.117                           & 0.554                           & 0.529                          & 0.141                                    \\
                                                                                 & StoryForest                    & 0.696                          & 0.725                                    & 0.592                           & 0.673                          & 0.765                          & 0.523                           & 0.697                           & 0.798                          & 0.596                                   \\
                                                                                 & Miranda                        & 0.706                          & 0.726                                    & 0.572                           & 0.694                          & 0.786                          & 0.571                           & 0.698                           & 0.791                          & 0.574                                    \\
                                                                                 & Staykovski                     & 0.669                          & 0.602                                    & 0.358                           & 0.697                          & 0.759                          & 0.487                           & 0.701                           & 0.765                          & 0.487                                   \\ \hdashline
\multirow{4}{*}{
\begin{tabular}[c]{@{}c@{}} PSE-variants of \\ the top baselines \end{tabular}} & Miranda-SenT5                  & 0.732                          & 0.753                                    & 0.617                           & 0.710                          & 0.798                          & 0.629                           & 0.717                           & 0.805                          & 0.644                                               \\
                                                                                 & Staykovski-SenT5               & 0.684                          & 0.631                                    & 0.415                           & 0.735                          & 0.798                          & 0.582                           & 0.704                           & 0.782                          & 0.537                                    \\
                                                                                 & Miranda-SenRB                  & 0.764                          & 0.785                                    & 0.648                           & 0.751                          & 0.835                          & 0.656                           & 0.759                           & 0.837                          & 0.657                                   \\
                                                                                 & Staykovski-SenRB               & 0.750                          & 0.720                                    & 0.567                           & 0.754                          & 0.824                          & 0.642                           & 0.762                           & 0.830                          & 0.660                                     \\ \hline
\multirow{6}{*}{Proposed}                                                        & \textbf{\algname{}-SenT5}      & 0.751$^*$                      & 0.763$^*$                                & 0.638$^*$                       & 0.780$^*$                      & 0.846$^*$                      & 0.694$^*$                       & 0.799$^*$                       & 0.861$^*$                      & 0.733$^*$                         \\
                                                                                 & \quad over baselines           & {\small $\triangle44.4\%$}     & {\small $\triangle136.0\%$}              & {\small $\triangle330.1\%$}     & {\small $\triangle35.9\%$}     & {\small $\triangle38.6\%$}     & {\small $\triangle160.5\%$}     & {\small $\triangle37.4\%$}      & {\small $\triangle33.0\%$}     & {\small $\triangle134.4\%$}      \\
                                                                                 & \quad over baslines-SenT5      & {\small $\blacktriangle6.2\%$} & {\small $\blacktriangle11.1\%$}          & {\small $\blacktriangle28.6\%$} & {\small $\blacktriangle8.0\%$} & {\small $\blacktriangle6.0\%$} & {\small $\blacktriangle14.8\%$} & {\small $\blacktriangle12.5\%$} & {\small $\blacktriangle8.5\%$} & {\small $\blacktriangle25.2\%$}   \\
                                                                                 & \textbf{\algname{}-SenRB}      & \textbf{0.789}$^*$             & \textbf{0.812}$^*$                       & \textbf{0.699}$^*$              & \textbf{0.810}$^*$             & \textbf{0.871}$^*$             & \textbf{0.739}$^*$              & \textbf{0.825}$^*$              & \textbf{0.880}$^*$             & \textbf{0.765}$^*$                \\
                                                                                 & \quad over baselines           & {\small $\triangle51.7\%$}     & {\small $\triangle151.2\%$}              & {\small $\triangle371.3\%$}     & {\small $\triangle41.1\%$}     & {\small $\triangle42.7\%$}     & {\small $\triangle177.4\%$}     & {\small $\triangle41.9\%$}      & {\small $\triangle36.0\%$}     & {\small $\triangle144.6\%$}       \\
                                                                                 & \quad over baselines-SenRB      & {\small $\blacktriangle4.2\%$} & {\small $\blacktriangle8.1\%$}           & {\small $\blacktriangle15.6\%$} & {\small $\blacktriangle7.6\%$} & {\small $\blacktriangle5.0\%$} & {\small $\blacktriangle13.9\%$} & {\small $\blacktriangle8.5\%$}  & {\small $\blacktriangle5.6\%$} & {\small $\blacktriangle16.2\%$}   \\ \hdashline
\multirow{3}{*}{
\begin{tabular}[c]{@{}c@{}} Variants of \\  \algname{}-SenRB \end{tabular}}      & \textbf{\quad w/o time-aware}  & 0.780                          & 0.801                                    & 0.682                           & 0.801                          & 0.864                          & 0.736                           & 0.817                           & 0.875                          & 0.760                               \\
                                                                                 & \textbf{\quad w/o theme-aware} & 0.767                          & 0.792                                    & 0.665                           & 0.771                          & 0.842                          & 0.679                           & 0.790                           & 0.856                          & 0.703                               \\
                                                                                 & \textbf{\quad w/o both}        & 0.752                          & 0.777                                    & 0.651                           & 0.753                          & 0.826                          & 0.677                           & 0.770                           & 0.843                          & 0.709                               \\
\bottomrule
\end{tabular}
    \vspace*{-0.0cm}
    \begin{tablenotes}
      \small
      \item $^*$ denotes statistically significant improvement (p<0.05 with the t-test) over the compared algorithms and $\triangle, \blacktriangle$ indicate the average improvement ratio.
    \end{tablenotes}
  \end{threeparttable}
\end{table*}

\noindent\textbf{\uline{Compared Algorithms.}} We compared \algname{} with the five online story discovery algorithms: \emph{ConStream}\cite{constream}, \emph{NewsLens}\,\cite{newslens}, \emph{StoryForest}\,\cite{storyforest}, \emph{Miranda}\,\cite{miranda}, and \emph{Staykovski}\,\cite{staykovski}\footnote{StoryForest and Miranda are adopted for an unsupervised setting.}. For a fair comparison, we also prepared their PSE-variants by adopting a PSE for embedding articles by averaging their sentence representations. We used two PSEs for \algname{} and the variants of baselines: Sentence-BERT\,\cite{sentencebert}\footnote{https://huggingface.co/sentence-transformers/all-roberta-large-v1} (i.e., AlgName-SenRB) and Sentence-T5\,\cite{sentencet5}\footnote{https://huggingface.co/sentence-transformers/sentence-t5-large}(i.e., AlgName-SenT5). Please note that for \algname{} and the PSE-variants of baselines, Sentence-BERT (i.e., -SenRB) is used as a default PSE unless otherwise specified.

\noindent\textbf{\uline{Implementation Details.}} All compared algorithms were implemented with \emph{Python 3.8.8} and evaluated on a Linux server with AMD EPYC 7502 32-Core CPU and 1TB RAM. We tokenized sentences and terms in articles by \emph{SpaCy 3.2.0} with \emph{en\_core\_web\_lg} pipeline\,\cite{en_core_web_lg} and counted (1,2)-gram terms by \emph{Sikit-learn 0.24.2}. For each algorithm, we used the default hyperparameters following the original work or tuned them to get the best results; specifically, the keywords size $N \in [1,15]$ and the temperature $T \in [1,5]$ for \algname{} ($N=10$ and $T=2$ were used by default); the standard score threshold for the micro-cluster assignment $\in [0,3]$ for \emph{ConStream}; the number of overlapping keywords for creating an edge $\in [1,5]$ and the similarity threshold for merging communities $\in [0,1]$ for \emph{NewsLens}; the number of keywords for an article $\in [1,20]$ to form a keyword graph for \emph{StoryForest}; the similarity threshold for cluster assignment $\in [1,5]$ for \emph{Miranda}; and the similarity threshold for creating an edge $\in [0, 0.5]$ and that for merging clusters $\in [0.5, 1]$ for \emph{Staykovski}. The minimum number $M$ of articles to form a valid story was set to the average number of articles in a story in a day, which was 8 for Newsfeed and 18 for WCEP18 and WCEP19.

\noindent\textbf{\uline{Evaluation Metrics}:} We used B-cubed F1 score\,(\emph{$B^3$-F1})\,\cite{b-cubed} for evaluating article-wise cluster quality and used Adjusted Mutual Information\,(\emph{AMI})\,\cite{ami} and Adjusted Rand Index\,(\emph{ARI})\,\cite{ari} for evaluating the mutual information and similarity of clustering results, respectively, adjusted for chance. For each metric, the average score over all sliding windows is reported to evaluate the overall performances over the entire news streams.


\smallskip
\subsection{Story Discovery Accuracy}
\label{sec:clustering_evaluation}
Table \ref{tbl:overall_accuracy} shows the overall evaluation results of all algorithms.

\noindent\textbf{\uline{Comparison with Baselines.}} 
\algname{} achieved higher $B^3$-F1, AMI, and ARI scores than baselines for all cases. For instance, \algname{}-SenRB achieved $44.9\%$ higher $B^3$-F1, $76.6\%$ higher AMI, and $231.1\%$ higher ARI than baselines when averaged over all cases. We also compared the PSE-variants of the top two baselines, Miranda and Staykovski (note that the PSE-variants of the other algorithms consistently performed worse than them). The PSE-variants improved all scores from their original versions, which shows the generic embedding power of a PSE. \algname{}, however, still outperformed them consistently in all cases for both SenT5 and SenRB. Specifically, \algname{}-SenRB achieved $6.8\%$ higher $B^3$-F1, $6.2\%$ higher AMI, and $15.2\%$ higher ARI than the SenRB-variants of the top two baselines when averaged over all cases. This clearly shows that \algname{} exploits a PSE more effectively with the help of thematic embedding.

\noindent\textbf{\uline{Ablation Study of \algname{}.}} We verified the efficacy of the theme and time-aware components employed in \algname{} by preparing the three variants: 
\vspace{-0.1cm}
\begin{itemize}[leftmargin=12pt, noitemsep]
\item \emph{w/o time-aware} does not consider the recency in thematic keywords and the time relevance in embedding articles and stories.
\item \emph{w/o theme-aware} does not consider the theme relevance in embedding articles and stories.
\item \emph{w/o both} does not consider both the time- and theme-aware components described above.
\end{itemize}


\noindent \algname{} consistently achieved higher scores than the above three variants, which indicates that both components jointly help to discover stories. More specifically, the theme-aware component contributed more than the time-aware component to the performance. These results show that temporally close articles in a story do not necessarily share thematically similar content only, so it is more critical to identify and filter out theme-irrelevant parts in each article for online story discovery. 



\begin{figure}[!t]
    \centering
    \includegraphics[width=\columnwidth]{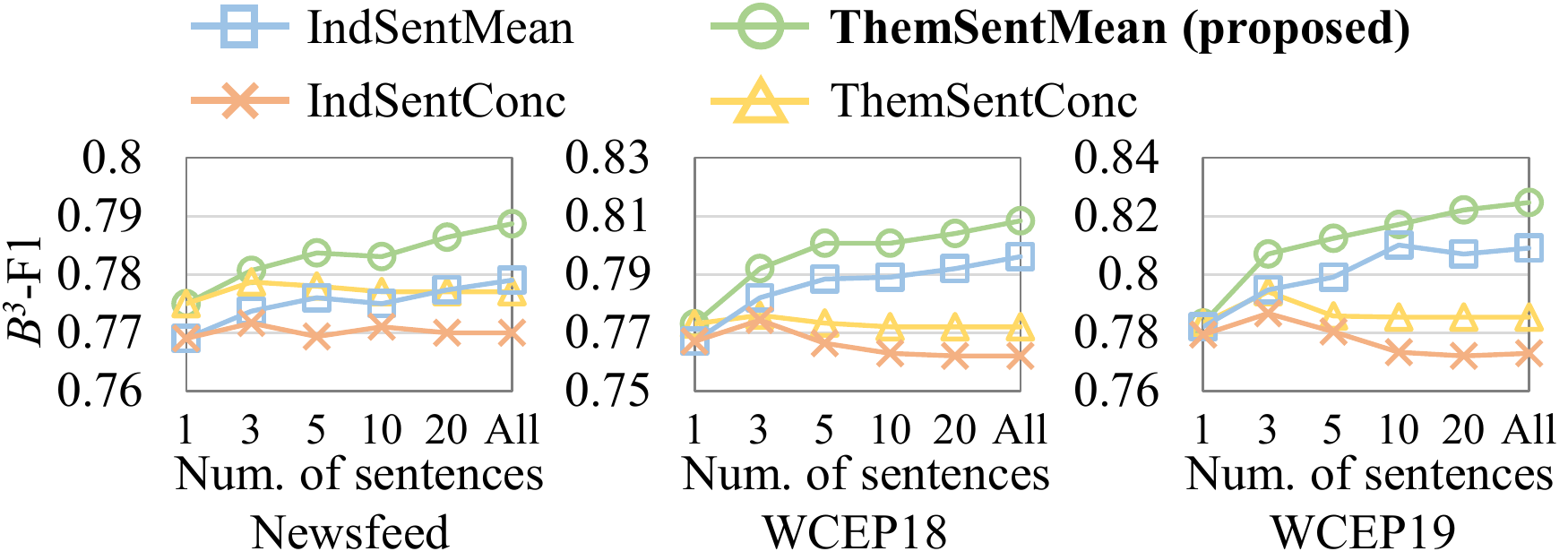}
    \vspace{-0.5cm}
    \caption{$B^3$-F1 scores of \algname{} with four strategies.}
    \vspace{-0.3cm}
    \label{fig:embedding_strategy}
\end{figure}

\begin{figure}[t]
    \centering
    \includegraphics[width=\columnwidth]{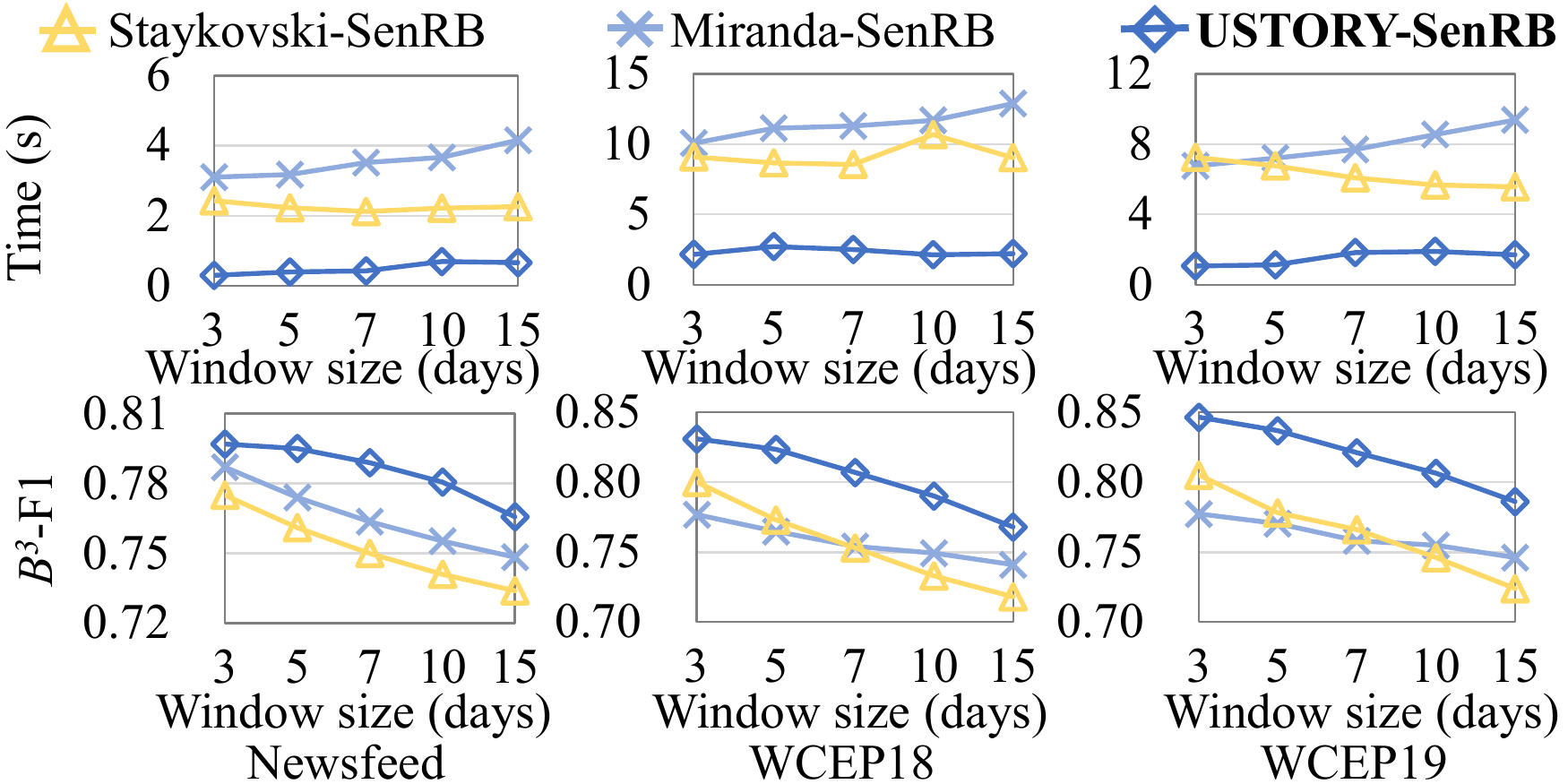}
    \vspace{-0.5cm}
    \caption{Varying sliding window sizes.}
    \vspace{-0.15cm}
    \label{fig:varying_W}      
\end{figure}

\smallskip
\subsection{Embedding Strategies Comparison}
\label{exp:alternatives}
We evaluated the story discovery performance of \algname{} with four embedding strategies to exploit a PSE: 
\begin{itemize}[leftmargin=12pt, noitemsep]
\item \emph{IndSentMean}: mean pooling of sentence representations
\item \emph{IndSentConc}: a representation of concatenated sentences
\item \emph{ThemSentMean} (default in \algname{}): weighted pooling of thematically prioritized sentence representations
\item \emph{ThemSentConc}: a representation of the concatenation of thematically prioritized sentences. 
\end{itemize}
Note that \algname{} automatically identifies temporal themes of news streams to prioritize sentences for ThemSentMean and ThemSentConc. We also varied the number of sentences used for embedding articles for further understanding of the compared strategies. Figure \ref{fig:embedding_strategy} shows the $B^3$-F1 results, while the results of AMI and ARI showed similar trends. The proposed ThemSentMean strategy in \algname{} showed the highest scores throughout the varying number of sentences. This indicates that thematic embedding effectively exploits all information in the article to discover stories. In general, the Mean strategies, IndSentMean and TheSentMean, outperformed the Concatenate strategies, IndSentConc and ThemSentConc, because the concatenated sentences can not incorporate all information in the article due to the input sequence limit of a PSE and also fail to capture local semantics. Meanwhile, it is notable that ThemSentConc still took advantage of the thematic prioritization of sentences and resulted in higher performance than IndSentConc.

\smallskip
\subsection{Scalability and Sensitivity Analysis}
\label{sec:hyperparameter}
\noindent\textbf{\uline{Scalability of \algname{}.}} We analyzed the scalability of \algname{} and the top two baselines on varying window sizes\,($W$), i.e., the number of articles in a window, which is the most critical factor affecting scalability. Figure \ref{fig:varying_W} shows the average wall clock time for the compared algorithms to process each sliding window and their $B^3$-F1 results. Note that AMI and ARI results showed similar trends. While the top two baselines resulted in higher processing time due to the high computation cost for community detection (Staykovski) and clustering (Miranda), \algname{} took a much shorter processing time for each sliding window. The increasing rate of processing time over larger windows for \algname{} was lower than Miranda and comparable to Staykovski in most cases. \algname{} also consistently achieve higher $B^3$-F1 scores than the other algorithms. This demonstrates the efficacy of story summary (PSS) for efficient and accurate story discovery.

\noindent\textbf{\uline{Processing Time Breakdown of \algname{}.}} We further broke down the processing time of \algname{} into the four main steps in Section \ref{sec:novelty_aware_clustering}. As shown in Figure \ref{fig:breakdown}, the initial article embedding step took the most processing time, followed by the story assignment step, the story summary (PSS) update step, and the seed stories discovery step. To further improve the scalability of \algname{}, especially for time-critical applications, the initial embedding step can be alternatively substituted with the IndSentMean strategy discussed in Section \ref{exp:alternatives}, which are expected to be faster but still effective for initialization purposes. Meanwhile, the last three steps in Newsfeed accounted for more portions of the total processing time compared with their proportions in the other data sets. This also conforms to the complexity analysis result, as Newsfeed has more concurrent stories in sliding windows.

\begin{figure}[!t]
    \centering
    \includegraphics[width=\columnwidth]{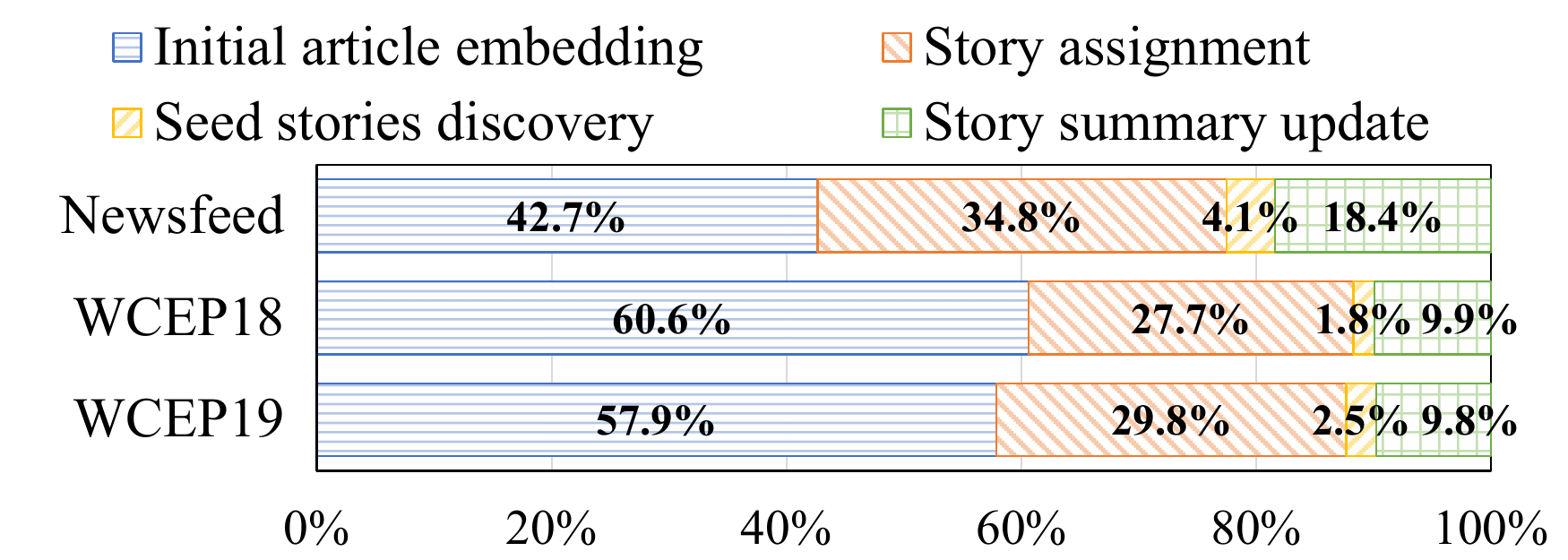}
    \vspace{-0.5cm}
    \caption{Processing time breakdown of \algname{}.}
    \vspace{-0.3cm}
    \label{fig:breakdown}
\end{figure}

\begin{figure}[t]
    \centering
    \includegraphics[width=\columnwidth]{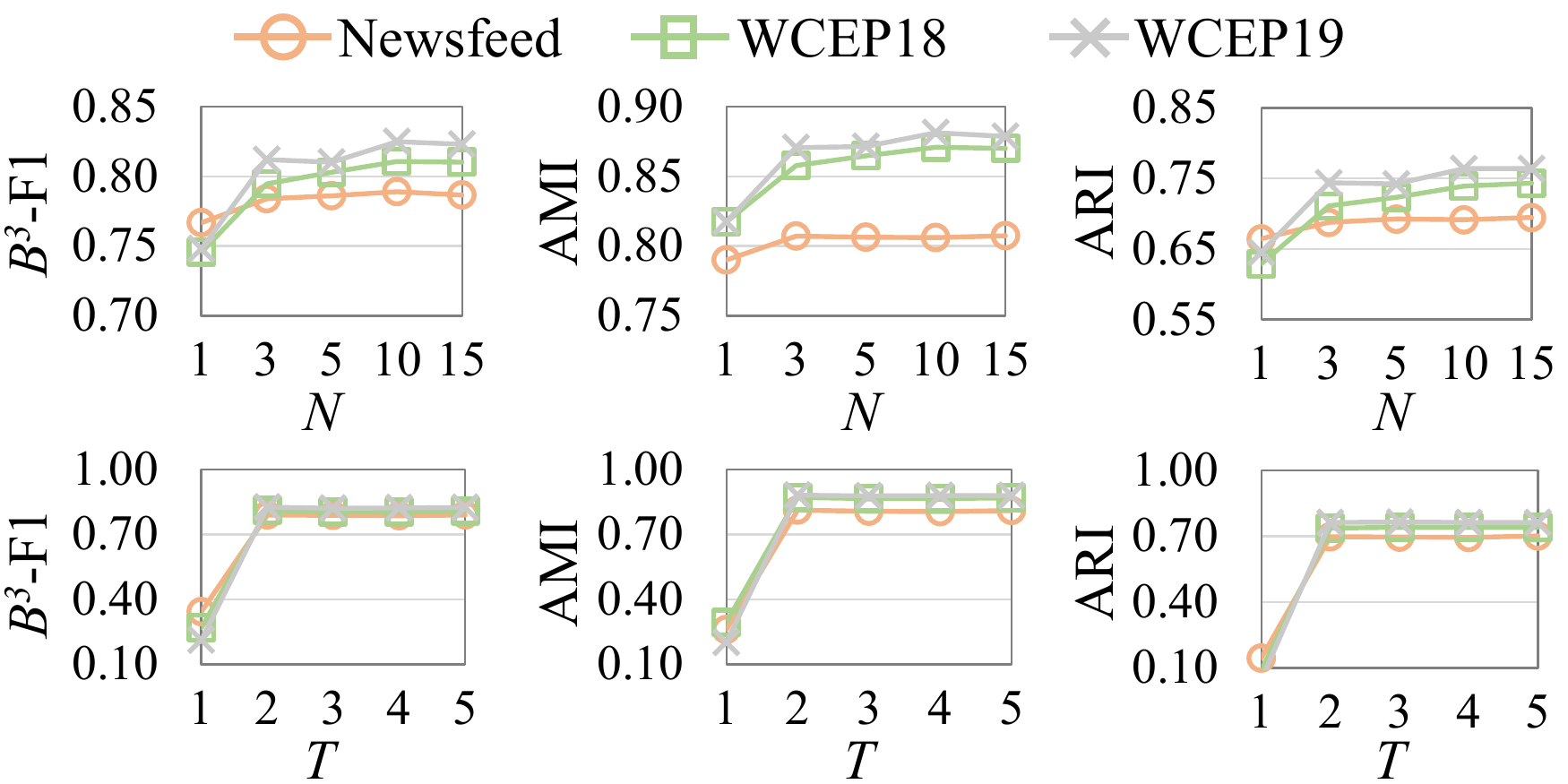}
    \vspace{-0.5cm}
    \caption{Varying keywords size $N$ and temperature $T$.}
    \vspace{-0.1cm}
    \label{fig:varying_TN}
\end{figure}

\smallskip
\noindent\textbf{\uline{Sensitivity of \algname{}.}} Figure \ref{fig:varying_TN} shows the sensitivity of \algname{} on the two main hyperparameters, keywords size $N$ and the temperature $T$, for the three performance metrics. The performances of \algname{} converge early at some points and become near optimal around the default values (i.e., $N=10$ and $T=2$). These trends show that setting the adequate number of keywords for identifying temporal themes (e.g., $N \!\geq\! 3$) and flattening the article-story confidence score distribution to at least some degree (e.g., $T \geq 2$) can lead to a robust story discovery performance.

\begin{figure*}[!t]
    \centering
    \text{\quad\quad\quad\quad\quad\quad\quad Titles, keywords, and sizes of discovered stories over time. \hspace{1.2cm} ThemSentMean. \quad\quad\quad\quad\quad IndSentMean.}    
    \includegraphics[width=0.83\paperwidth, trim= 0 0.75cm 0 0, clip]{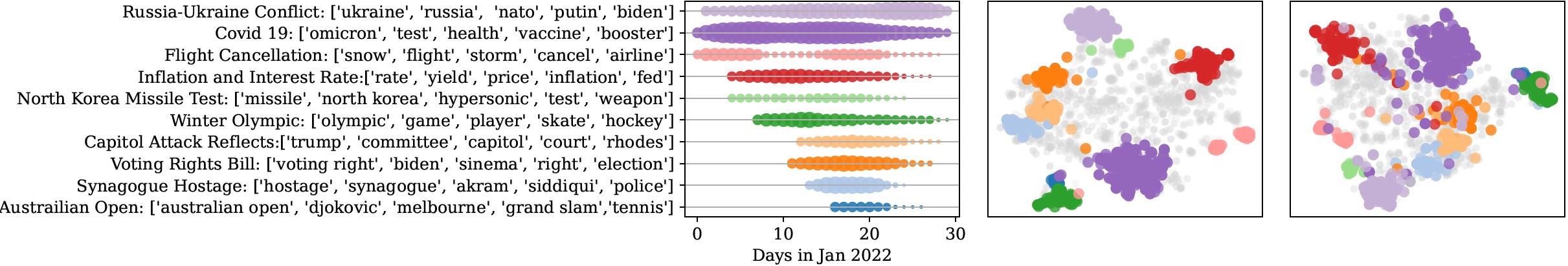}
    \vspace{-0.8cm}
    \caption{The timeline and visualizations of stories discovered by \algname{} in USNews (best viewed in color).}
    \label{fig:casestudy}
    \vspace{-0.2cm}    
\end{figure*}

\begin{figure}[!t]
    \centering
    \includegraphics[width=\columnwidth]{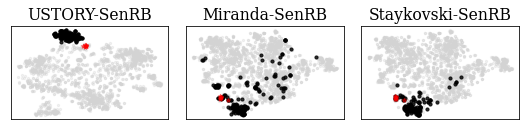} 
    \vspace{-0.8cm}
    \caption{Visualizations of the clusters about \textsf{Russia-Ukraine Conflict} (in black circles) and \textsf{North Korea Missile Test} (mis-clustered by the other algorithms, in red stars).}
    \label{fig:russia_ukraine}
    \vspace{-0.56cm}
\end{figure}

\newpage
\subsection{Case Study}
\label{sec:casestudy}
We conducted a qualitative case study on USNews with \algname{} and the PSE-variants of the top two baselines. Figure \ref{fig:casestudy} visualizes the ten discovered stories and their article embeddings with \algname{} and the PSE-variants of baselines. The leftmost sub-figure shows a timeline of the sizes of stories with their titles (manually given by the authors for convenience) and their top five thematic keywords (automatically identified by \algname{}). 

\algname{} successfully identified and tracked the important real news stories. For instance, the long-term stories such as \textsf{Russia-Ukraine Conflict} and \textsf{Covid 19}, which have been constant issues in recent years, were continuously discovered by \algname{} throughout the whole period. At the same time, the short-term stories such as \textsf{North Korea Missile Test} and \textsf{Capitol Attack Reflects} that have been discussed for some days were also timely discovered by \algname{}. The resulting embedding space also shows that the articles are distinctively clustered by their stories. Following the two embedding strategies discussed in Section \ref{exp:alternatives}, the right two sub-figures respectively visualizes the article representations embedded by ThemSentMean (used by default in \algname{}) and IndSentMean (used in the PSE-variants of baselines) in a 2D-space through t-SNE\,\cite{tsne}. As expected, the discovered stories are clearly distinguished with the ThemSentMean embedding, while they are more cluttered and mixed with the IndSentMean embedding.

We further analyzed the differences between the compared algorithms in detail by drilling down into a specific story, \textsf{Russia-Ukraine Conflict}. Figure \ref{fig:russia_ukraine} shows the 2D-visualizations of the articles in the clusters related to \textsf{Russia-Ukraine Conflict} discovered by each algorithm. It is clear that \algname{} discovered a more compact cluster of articles about \textsf{Russia-Ukraine Conflict}.
There are some articles about \textsf{North Korea Missile Test} mis-clustered together \emph{only by the PSE-variants of baselines}. This is because some articles about \textsf{North Korea Missile Test} contain general contents on war, weapons, or conflicts, which are also frequently found in the articles about \textsf{Russia-Ukraine Conflict}. However, \algname{} successfully distinguished the two stories in the embedding space by focusing on only the theme-relevant parts of articles; the example prioritized sentences are given in Table \ref{tbl:russia_ukraine}. 


\begin{table}[!t]
    \caption{The top-3 and bottom-3 sentences prioritized by \algname{} in an example article about \textsf{Russia-Ukraine Conflict}.}
    \vspace{-0.3cm}
    \label{tbl:russia_ukraine}
    \small
\begin{tabular}{m{0.2cm}L{7.5cm}}
\toprule
s\#  & Sentence \\ \midrule                      
s04   & President Joe Biden has warned that Russia will face severe economic consequences if Russian President Vladimir Putin were to launch an invasion of Ukraine. \\
s15 & Russia says it feels threatened by the prospect of the U.S. deploying offensive missile systems in Ukraine, even though Biden has assured Putin he has no intention of doing so. \\
s01 & U.S. suggests it's open to limiting military exercises, missile deployments in talks with Russia. \\
\ldots & \ldots \\
s08 & The Geneva talks, to be followed by other sessions next week in Brussels and Vienna, are aimed at averting a crisis. \\
s14 & "Both sides would need to make essentially the same commitment." \\
s27 & These actions could also restrict export of products made abroad if they contain more than a specified percentage of U.S. content. \\
\bottomrule
\end{tabular}

    \vspace*{-0.06cm}
\end{table}

\section{Discussion and conclusion}
While we followed an unsupervised approach in this work, it would be promising to extend the idea of thematic embedding in a weakly supervised setting by assuming that story-indicative external knowledge is available (e.g., category\,\cite{lotclass}, entity\,\cite{saravan}, taxonomy\,\cite{topicexpan}, or timeline\,\cite{retro2}). They can auxiliarily help identify temporal themes and also allow a more flexible time scope of ongoing stories; \algname{} can refer to an extra set of landmark stories with their lightweight summaries to capture sparse stories with longer timelines.

In conclusion, this paper proposed \algname{}, a scalable framework for unsupervised online story discovery from news streams. To deal with text-rich and evolving news streams, \algname{} employs a novel idea of thematic embedding with a pretrained sentence encoder and conducts novelty-aware adaptive clustering by managing compact story summaries. Extensive experiments on real news data sets showed that \algname{} outperforms the existing baselines and their variants, while being scalable and robust to various experiment settings. We believe that \algname{} can be used for many downstream real-time applications to help people understand unstructured and massive new streams.




\begin{acks}
The first author was supported by the National Research Foundation of Korea (Basic Science Research Program: 2021R1A6A3A14043765). The research was supported in part by US DARPA (FA8750-19-2-1004 and HR001121C0165), National Science Foundation (IIS-19-56151, IIS-17-41317, IIS 17-04532, 2019897, and 2118329), and the Institute of Information and Communications Technology Planning
and Evaluation (IITP) in Korea (2020-0-01361).
\end{acks}

\clearpage



\bibliographystyle{ACM-Reference-Format}

\balance
\bibliography{reference}

\end{document}